\documentclass[]{aastex631}

\defcitealias{Gordon2023}{G23}

\usepackage{amsmath}	% Advanced maths commands
\usepackage{amssymb}	% Extra maths symbols
\usepackage{subfigure}

\received{November 14, 2023}
\accepted{\today}

\submitjournal{RNAAS}

\shorttitle{63 New Giant Radio Galaxies in FIRST}
\shortauthors{Ramdhanie et al.}
\begin{document}

\title{The Discovery of 63 Giant Radio Galaxies in the FIRST Survey}

\correspondingauthor{Yjan~A. Gordon}
\email{yjan.gordon@wisc.edu}

\author{Soren Ramdhanie}
\affiliation{Department of Astronomy, University of Wisconsin-Madison, 475 N. Charter St., Madison, WI 53703, USA}
\affiliation{Department of Physics, University of the West Indies, St. Augustine, Trinidad and Tobago}
\altaffiliation{Soren Ramdhanie and Brianna Sampson are summer students at the National Radio Astronomy Observatory.}

\author[0000-0003-1432-253X]{Yjan~A. Gordon}
\affil{Department of Physics, University of Wisconsin-Madison, 
1150 University Ave, Madison, WI 53706, USA}

%%%rest alphabetical

\author[0000-0003-4873-1681]{Heinz Andernach}
\affiliation{Th\"uringer Landessternwarte, Sternwarte 5, D-07778 Tautenburg, Germany}
\affiliation{Permanent Address: Depto. de Astronom\'ia, Univ. de Guanajuato,
Callej\'on de Jalisco s/n, Guanajuato, C.P. 36023, GTO, Mexico} %%(if heinz checks his list and wants a coauthorship)

\author[0000-0003-0713-3300]{Eric~J. Hooper}
\affiliation{Department of Astronomy, University of Wisconsin-Madison, 
475 N. Charter St., Madison, WI 53703, USA}

\author{Brianna Sampson}
\affiliation{Department of Astronomy, University of Wisconsin-Madison, 475 N. Charter St., Madison, WI 53703, USA}
\affiliation{Department of Physics, University of the West Indies, St. Augustine, Trinidad and Tobago}
\altaffiliation{Soren Ramdhanie and Brianna Sampson are summer students at the National Radio Astronomy Observatory.}



\begin{abstract}

Giant Radio Galaxies (GRGs) are Active Galactic Nuclei (AGN) with radio emission that extends over projected sizes $>0.7\,$Mpc.
The large angular sizes associated with GRGs complicate their identification in radio survey images using traditional source finders.
In this Note, we use \textsc{DRAGNhunter}, an algorithm designed to find double-lobed radio galaxies, to search for GRGs in the Faint Images of the Radio Sky at Twenty cm survey (FIRST). 
Radio and optical images of identified candidates are visually inspected to confirm their authenticity, resulting
in the discovery of $63$ previously unreported GRGs.
% in $80$ GRGs being found.
% Comparisons with the literature show that $17$ of these GRGs have been previously identified, with the remaining $63$ being new discoveries.
%% abstract length = 91 words

\end{abstract}

\keywords{\href{http://astrothesaurus.org/uat/654}{Giant radio galaxies (654)},
\href{http://astrothesaurus.org/uat/1343}{Radio Galaxies (1343)},
\href{http://astrothesaurus.org/uat/508}{Extragalactic radio sources (508),
\href{http://astrothesaurus.org/uat/16}{Active galactic nuclei (16)}}
}

%%%word limit = 1500 include title, captions, headers, references

%%%%1930 words at present.
%%%% 1456 words as of Nov 9th (thanks Heinz and Eric!!!)

%%%%%%%%%%%%%%%%%%%%%%%%%%%%%%%%%%%
%%%%%%%%%%%%%%%%%%%%%%%%%%%%%%%%%%%
\section{Introduction} 
\label{sec:intro}

Radio galaxies with a projected ``Largest Linear Size'' (LLS) over $700\,$kpc are considered Giant Radio Galaxies \citep[GRGs, e.g.,][]
{Andernach2021}.
Their jets should be intrinsically powerful \citep{Wiita1986} and relatively uninhibited \citep{Dabhade2020SAGAN} to reach such sizes, and can provide insights into the AGN central engine and host galaxy environment.
GRGs can be difficult to identify in radio surveys: for a classic Double Radio Source associated with an Active Galactic Nucleus \citep[DRAGN,][]{Leahy1993}, the lobes of the radio galaxy may be detected as multiple sources despite belonging to the same physical object.
Consequently, many detectable GRGs may remain unidentified.

Recently, \citet[][hereafter \citetalias{Gordon2023}]{Gordon2023} introduced \textsc{DRAGNhunter}, an algorithm that groups `extended components' into physical sources, and applied it to the Very Large Array Sky Survey \citep[VLASS,][]{Lacy2020}, discovering 31 new GRGs in the process.  
Given the lower frequency ($1.4\,$GHz) and better sensitivity to extended sources of FIRST \citep{Becker1995}, we used \textsc{DRAGNhunter} on the FIRST catalog \citep{Helfand2015} to identify further unidentified GRGs.

\vspace{1em}

%%%%%%%%%%%%%%%%%%%%%%%%%%%%%%%%%%%
%%%%%%%%%%%%%%%%%%%%%%%%%%%%%%%%%%%
\section{Identifying Candidate GRGs}
\label{sec:candidate_GRGs}

\textsc{DRAGNhunter} identifies DRAGNs by pairing cataloged extended radio sources based on their separation and relative alignment, then uses the likelihood ratio approach \citep{McAlpine2012} to search for the probable host galaxy in the AllWISE catalog \citep{Cutri2014}.
Hosts are then cross-matched with a number of legacy catalogs to obtain redshifts \citepalias[for details see Sections 2-4 of][]{Gordon2023}.
Selecting extended FIRST sources as having deconvolved sizes $ >5.5''$, \textsc{DRAGNhunter} finds  $>9,000$ DRAGNs with a probable host and redshift.
We determine their LLS from their measured Largest Angular Size (LAS) assuming a flat $\Lambda$CDM cosmology with $H_{0} = 70\,$km/s/Mpc, $\Omega_{\text{M}} = 0.3$,
and \text{$\Omega_{\Lambda} = 0.7$.} 
We obtain $\text{LLS} > 0.7\,$Mpc for 213 DRAGNs, but because \textsc{DRAGNhunter} is an automated algorithm, we inspect
their radio and optical images to exclude any mistaken groupings of 
independent sources, and ensure that the identified host and redshift are correct.
% Of note is one particular case where the AllWISE host, \text{WISEA 
% J010930.98-023722.0}, is a blend of three potential optical hosts,
% all at similar redshifts.
% Here we adopt the most central of these as our host.
Furthermore, we manually remeasure the LAS that \textsc{DRAGNhunter} automatically estimates, ensuring these truly are giants.
In total this leaves us with $80$ GRGs identified by \textsc{DRAGNhunter} in FIRST.

\vspace{1em}

%%%%%%%%%%%%%%%%%%%%%%%%%%%%%%%%%%%
%%%%%%%%%%%%%%%%%%%%%%%%%%%%%%%%%%%
\section{Newly Discovered Giants}
\label{sec:results}

We cross match these GRGs with a comprehensive list of the $>3,000$ known GRGs previously reported by: 
\citet{Kuzmicz2018}, \citet{Koziel-Wierzbowska2020}, \citet{Dabhade2020SAGAN}, \citet{Dabhade2020LoTSS}, \citet{Galvin2020}, 
\citet{Kuzmicz2021}, \citet{Andernach2021}, \citet{Mahato2022}, \citet{Oei2023}, and \citetalias{Gordon2023}.
Of our GRGs, $17$ are within $5''$ of known GRGs and are not new discoveries.
We list the remaining $63$ newly discovered GRGs in Table \ref{tab:grgs}.

Only $6$ of our GRGs have spectroscopic redshifts, all from the Sloan Digital Sky Survey \citep[SDSS-DR16,][]{Ahumada2020}.
The remaining GRGs have photometric redshifts (photo-zs) obtained from the Dark Energy Spectroscopic Instrument 
imaging Legacy Surveys \citep[LS-DR8,][]{Duncan2022}.  
Alternative photo-zs are available for many of our GRGs, and the choice of photo-z may impact the GRG classification for sources with \text{$\text{LLS}\approx 0.7\,$Mpc}. 
However, \citet{Duncan2022} test the robustness of their photo-zs for radio galaxies, finding them to be reliable over the redshift range of our sample, thus these are an appropriate choice here.

Although none of our $63$ sources have been reported as GRGs, they may have been identified as extended radio galaxies in previous catalogs 
\citep[e.g.,][]{Amirkhanyan2009, Proctor2011}.
For example, we measure $\text{LAS} = 96''$ for \text{SDSS J025551.47-025939.2},
whereas \citet{Kuzmicz2021} 
only measure $80''$ leading to $\text{LLS}=0.64\,$Mpc in their work. 
Here, diffuse structure in the northern lobe extends beyond the hotspot used to measure the LAS by \citet{Kuzmicz2021}.
Finally, we note that \textsc{DRAGNhunter} is biased toward finding relatively unbent and symmetric \text{FR II} 
galaxies \citepalias{Gordon2023}.
Consequently, GRGs with atypical morphologies will likely be missed. 
Nonetheless, this Note demonstrates the usefulness of automated routines in finding GRGs. 

%%%%%%%%%%%%%%%%%%%%%%%%%%%%%%%%%%%%%%%%%%%%%%%%%%%%%%%
%%%%%%%%%%%data table

%for arxiv version include full table at end (eg table for RN)
%\newpage 
\startlongtable
\begin{deluxetable*}{rccccccccc}
    \tablecaption{New GRGs in FIRST
    \label{tab:grgs}}
    % \tabletypesize{\footnotesize}
    \tablehead{\colhead{Name} & \colhead{RAJ2000} & \colhead{DEJ2000} & \colhead{rmag} & \colhead{$z$} & \colhead{$z_{\text{type}}$} & \colhead{$S_{\text{FIRST}}$} & \colhead{$\log_{10}L_{\text{FIRST}}$} & \colhead{LAS} & \colhead{LLS}\\
    \colhead{ } & \colhead{deg} & \colhead{deg} & \colhead{mag} & \colhead{ } & \colhead{ } & \colhead{mJy} & \colhead{$\text{W}\,\text{Hz}^{-1}$} & \colhead{arcsec} & \colhead{Mpc}}
    \decimalcolnumbers
    \startdata
        SDSS J000133.55$-$105738.0 & 0.3898 & -10.9606 & 20.06 & 0.675 & p & 43.60 & 26.09 & 115 & 0.81 \\
        SDSS J001217.46$+$120733.5 & 3.0728 & 12.1260 & 21.93 & 0.913 & p & 42.46 & 26.45 & 101 & 0.79 \\
        SDSS J001403.56$+$053008.1 & 3.5148 & 5.5023 & 19.82 & 0.51571 & s & 15.72 & 25.34 & 135 & 0.84 \\
        DELS J003912.82$+$002319.6 & 9.8034 & 0.3888 & 24.50 & 1.261 & p & 44.06 & 26.86 & 85 & 0.71 \\
        SDSS J005843.94$+$060352.7 & 14.6831 & 6.0646 & 22.02 & 0.778 & p & 17.34 & 25.86 & 95 & 0.71 \\
        SDSS J010541.66$-$031554.3 & 16.4236 & -3.2650 & 22.39 & 0.988 & p & 69.28 & 26.75 & 91 & 0.73 \\
        SDSS J010931.12$-$023723.8 & 17.3797 & -2.6233 & 22.29 & 1.10045 & s & 15.54 & 26.24 & 114 & 0.93 \\
        SDSS J015350.13$+$023539.5 & 28.4589 & 2.5943 & 21.03 & 0.650 & p & 65.22 & 26.22 & 112 & 0.78 \\
        SDSS J015438.76$+$120212.3 & 28.6615 & 12.0368 & 21.85 & 0.83175 & s & 75.46 & 26.58 & 146 & 1.11 \\
        SDSS J020047.38$-$023629.0 & 30.1975 & -2.6081 & 21.14 & 0.748 & p & 43.10 & 26.21 & 145 & 1.06 \\
        SDSS J023303.66$+$012837.1 & 38.2653 & 1.4771 & 22.87 & 0.935 & p & 8.04 & 25.75 & 92 & 0.72 \\
        SDSS J025551.47$-$025939.2 & 43.9645 & -2.9942 & 17.74 & 0.96949 & s & 115.62 & 26.95 & 96 & 0.76 \\
        SDSS J031516.73$+$004906.3 & 48.8197 & 0.8185 & 21.99 & 0.962 & p & 55.91 & 26.63 & 95 & 0.75 \\
        SDSS J083504.48$+$092803.7 & 128.7687 & 9.4677 & 21.09 & 0.698 & p & 48.89 & 26.25 & 114 & 0.81 \\
        SDSS J085102.17$+$045844.1 & 132.7591 & 4.9789 & 21.89 & 0.692 & p & 25.99 & 25.90 & 122 & 0.87 \\
        DELS J091150.56$-$072648.8 & 137.9607 & -7.4469 & 23.35 & 1.049 & p & 36.94 & 26.55 & 93 & 0.75 \\
        DELS J092719.46$+$242027.0 & 141.8309 & 24.3407 & 22.01 & 0.589 & p & 52.22 & 26.01 & 130 & 0.86 \\
        DELS J093016.68$+$114241.4 & 142.5695 & 11.7115 & 22.77 & 1.147 & p & 21.59 & 26.43 & 147 & 1.21 \\
        DELS J094554.69$+$223020.1 & 146.4779 & 22.5056 & 23.98 & 1.091 & p & 34.05 & 26.57 & 137 & 1.12 \\
        SDSS J100302.65$+$174145.6 & 150.7611 & 17.6960 & 22.12 & 0.963 & p & 14.62 & 26.05 & 114 & 0.90 \\
        DELS J101056.04$+$023639.2 & 152.7335 & 2.6109 & 24.09 & 1.144 & p & 16.45 & 26.31 & 116 & 0.95 \\
        SDSS J101206.08$+$184956.0 & 153.0253 & 18.8322 & 20.75 & 0.673 & p & 34.86 & 25.99 & 104 & 0.73 \\
        SDSS J101247.79$+$132703.5 & 153.1991 & 13.4510 & 21.84 & 0.665 & p & 17.04 & 25.67 & 118 & 0.83 \\
        DELS J103604.61$+$201103.4 & 159.0192 & 20.1843 & 22.51 & 0.994 & p & 93.22 & 26.89 & 139 & 1.11 \\
        DELS J105706.88$+$225621.8 & 164.2787 & 22.9394 & 23.92 & 0.988 & p & 11.78 & 25.99 & 97 & 0.77 \\
        DELS J112456.90$+$235242.2 & 171.2371 & 23.8784 & 23.41 & 1.057 & p & 13.23 & 26.12 & 133 & 1.08 \\
        DELS J115635.66$+$081655.9 & 179.1486 & 8.2822 & 23.13 & 1.034 & p & 19.86 & 26.27 & 144 & 1.16 \\
        SDSS J121008.57$+$184602.5 & 182.5357 & 18.7673 & 21.49 & 0.857 & p & 53.53 & 26.47 & 129 & 0.99 \\
        SDSS J121128.84$+$145034.0 & 182.8702 & 14.8428 & 22.92 & 1.068 & p & 41.11 & 26.62 & 105 & 0.85 \\
        SDSS J121358.95$+$000650.8 & 183.4956 & 0.1141 & 21.21 & 0.770 & p & 126.77 & 26.71 & 124 & 0.92 \\
        DELS J121729.04$+$264126.1 & 184.3710 & 26.6906 & 24.94 & 1.260 & p & 13.70 & 26.14 & 102 & 0.85 \\
        DELS J122257.93$+$084449.9 & 185.7414 & 8.7472 & 24.22 & 1.302 & p & 13.75 & 26.39 & 91 & 0.76 \\
        SDSS J123831.55$+$305354.2 & 189.6315 & 30.8984 & 21.60 & 0.654 & p & 37.25 & 25.99 & 132 & 0.92 \\
        SDSS J124608.79$+$190625.5 & 191.5367 & 19.1071 & 22.71 & 1.059 & p & 28.03 & 26.45 & 130 & 1.05 \\
        DELS J125320.32$+$162448.2 & 193.3347 & 16.4134 & 23.48 & 0.909 & p & 16.01 & 26.02 & 107 & 0.84 \\
        DELS J130058.08$+$300713.4 & 195.2420 & 30.1204 & 22.95 & 0.881 & p & 46.82 & 26.44 & 97 & 0.75 \\
        DELS J130956.06$+$132316.0 & 197.4836 & 13.3878 & 21.40 & 0.833 & p & 21.79 & 26.04 & 114 & 0.87 \\
        DELS J131931.48$+$211230.6 & 199.8812 & 21.2085 & 22.70 & 0.979 & p & 20.67 & 25.98 & 104 & 0.83 \\
        SDSS J134357.80$+$221813.2 & 205.9909 & 22.3036 & 21.44 & 0.605 & p & 61.82 & 26.12 & 110 & 0.74 \\
        SDSS J135435.29$-$042027.3 & 208.6471 & -4.3409 & 20.78 & 0.819 & p & 14.25 & 25.84 & 124 & 0.94 \\
        DELS J135703.76$+$243614.0 & 209.2657 & 24.6039 & 22.58 & 0.973 & p & 22.42 & 26.25 & 100 & 0.80 \\
        SDSS J141042.41$+$155334.9 & 212.6767 & 15.8930 & 22.49 & 0.878 & p & 46.22 & 26.43 & 92 & 0.71 \\
        DELS J142640.10$+$133051.1 & 216.6671 & 13.5142 & 23.61 & 0.805 & p & 18.92 & 25.94 & 93 & 0.70 \\
        SDSS J144238.00$+$261326.7 & 220.6583 & 26.2241 & 22.12 & 0.938 & p & 23.79 & 26.23 & 92 & 0.73 \\
        SDSS J145528.62$+$104451.1 & 223.8693 & 10.7475 & 19.95 & 0.43883 & s & 42.76 & 25.59 & 149 & 0.85 \\
        SDSS J150041.03$-$021511.7 & 225.1710 & -2.2533 & 22.42 & 0.896 & p & 27.77 & 26.24 & 95 & 0.74 \\
        DELS J151353.56$+$135919.3 & 228.4732 & 13.9887 & 23.55 & 1.003 & p & 15.06 & 26.11 & 95 & 0.76 \\
        SDSS J151751.03$+$095936.8 & 229.4627 & 9.9936 & 20.86 & 0.689 & p & 337.85 & 27.01 & 104 & 0.74 \\
        SDSS J154114.86$+$300257.5 & 235.3119 & 30.0493 & 21.64 & 0.692 & p & 7.14 & 25.34 & 122 & 0.87 \\
        SDSS J154905.62$+$264331.4 & 237.2734 & 26.7254 & 20.89 & 0.779 & p & 188.68 & 26.90 & 105 & 0.78 \\
        SDSS J155254.45$+$312656.5 & 238.2269 & 31.4491 & 21.17 & 0.738 & p & 7.74 & 25.45 & 98 & 0.71 \\
        DELS J160431.82$+$304625.3 & 241.1326 & 30.7737 & 23.09 & 1.112 & p & 13.86 & 26.20 & 100 & 0.82 \\
        DELS J162417.37$+$153937.8 & 246.0724 & 15.6605 & 23.54 & 0.921 & p & 86.42 & 26.76 & 90 & 0.71 \\
        SDSS J163051.31$+$532518.7 & 247.7138 & 53.4219 & 21.23 & 0.78579 & s & 12.20 & 25.72 & 98 & 0.73 \\
        DELS J170158.23$+$225940.6 & 255.4925 & 22.9948 & 22.94 & 0.748 & p & 12.83 & 25.69 & 121 & 0.89 \\
        SDSS J210537.99$+$101851.7 & 316.4083 & 10.3144 & 22.02 & 0.607 & p & 76.47 & 26.21 & 121 & 0.81 \\
        SDSS J212953.43$-$002404.3 & 322.4727 & -0.4012 & 20.81 & 0.762 & p & 126.52 & 26.65 & 128 & 0.94 \\
        SDSS J221607.43$-$044752.4 & 334.0310 & -4.7979 & 22.38 & 0.756 & p & 18.95 & 25.87 & 118 & 0.87 \\
        DELS J225123.88$-$103459.1 & 342.8495 & -10.5831 & 22.15 & 1.011 & p & 132.26 & 27.06 & 113 & 0.91 \\
        DELS J225125.27$-$025451.8 & 342.8553 & -2.9144 & 20.19 & 1.324 & p & 7.37 & 26.14 & 110 & 0.92 \\
        DELS J232804.71$+$001933.4 & 352.0196 & 0.3260 & 24.54 & 1.195 & p & 12.42 & 26.24 & 104 & 0.86 \\
        DELS J234027.85$+$003057.4 & 355.1161 & 0.5160 & 23.60 & 1.010 & p & 139.88 & 27.09 & 87 & 0.70 \\
        SDSS J235527.42$+$010853.5 & 358.8643 & 1.1482 & 22.05 & 0.783 & p & 10.54 & 25.65 & 113 & 0.84\\
    \enddata
    \tablecomments{For these 63 GRGs, we list the host name (1), R.A. (2), declination (3), $r$-band magnitude from LS-DR8 (4), redshift (5), photometric (p) or spectroscopic (s) redshift (6), flux density in FIRST (7), $1.4\,$GHz luminosity (8), LAS (9) and LLS (10).
    The full table is available in machine readable format in the online article and from \href{https://doi.org/10.5281/zenodo.10094175}{https://doi.org/10.5281/zenodo.10094175}.}
\end{deluxetable*}

%%%%%%%%%%%%%%%%%%%%%%%%%%%%%%%%%%%%%%%%%%%%%%%%%%%%%%%%%%%%%%%%%%%%%%%%%%%%%%%%%%%%%%%%%%%%%%%%%%%

%% Also note that the akcnowlodgment environment does not support long amounts of text. If you have a lot of people and institutions to acknowledge, do not use this command. Instead, create a new \section{Acknowledgments}.
% \begin{acknowledgments}
\section*{}\noindent
S.R. and B.S. acknowledge support from the NSF under Cooperative Agreements No. 1647375 and 1647378, including the Radio Astronomy Data Imaging and Analysis Lab (RADIAL) Research \& Training Experience program. 
S.R. and B.S. further acknowledge support from the Alfred P. Sloan Foundation’s Creating Equitable Pathways to STEM Graduate Education program and NSF grant AST 21-50222.
% \end{acknowledgments}


% \vspace{5mm}
\facilities{CDS, VLA, WISE, Sloan}


\software{Astropy \citep{Astropy2022}, NumPy \citep{Harris2020}, SAO DS9 \citep{Joye2003}, Topcat \citep{Taylor2005}}



% \appendix

% \section{Appendix A}

% \pagebreak
% \newpage

\bibliography{GRGs_in_FIRST.bib}{}
\bibliographystyle{aasjournal}



\end{document}